# Efficient High Capacity Steganography Technique

Alan Anwer Abdulla, Sabah A. Jassim and Harin Sellahewa

University of Buckingham, Buckingham MK18 1EG, UK

alananwer@yahoo.com {sabah.jassim; harin.sellahewa}@buckingham.ac.uk

## ABSTRACT

Performance indicators characterizing modern steganographic techniques include capacity (i.e. the quantity of data that can be hidden in the cover medium), stego quality (i.e. artifacts visibility), security (i.e. undetectability), and strength or robustness (intended as the resistance against active attacks aimed to destroy the secret message). Fibonacci based embedding techniques have been researched and proposed in the literature to achieve efficient steganography in terms of capacity with respect to stego quality. In this paper, we investigated an innovative idea that extends Fibonacci-like steganography by bit-plane(s) mapping instead of bit-plane(s) replacement. Our proposed algorithm increases embedding capacity using bit-plane mapping to embed two bits of the secret message in three bits of a pixel of the cover, at the expense of a marginal loss in stego quality. While existing Fibonacci embedding algorithms do not use certain intensities of the cover for embedding due to the limitation imposed by the Zeckendorf theorem, our proposal solve this problem and make all intensity values candidates for embedding. Experimental results demonstrate that the proposed technique double the embedding capacity when compared to existing Fibonacci methods, and it is secure against statistical attacks such as RS, POV, and difference image histogram (DIH).

**Keywords:** Steganography, LSB, Fibonacci, bit-plane(s) mapping

## 1. INTRODUCTION

Steganography is a mechanism used to embed (conceal) a secret message within another message during seemingly mundane communication sessions in such a way that only the sender and intended recipient are aware of the existence of the secret message. In steganography, the secret information one wishes to send is called the message. This message is embedded in a cover, which is typically an image, a video or an audio file. After the message is embedded, the cover becomes a stego. In this paper the cover image has been used. The most popular and frequently used steganographic method is the Least Significant Bit embedding (LSB). LSB steganography is based on manipulating the LSB planes by directly replacing the LSB of the cover-image with the message bits. One of the well-known bitplane representations of pixel value is binary bitplane and almost all the steganographic methods are used binary representation. Once the pixel values of the cover image are converted into binary form, the secret message is going to embed by replacing it with the LSB. In the proposed method, Fibonacci bitplane representation has used. The purpose of using Fibonacci bitplane is explained in the next section. In grey level image, to represent pixel values in the range 0-255 in binary 8 bits are needed, while in Fibonacci 12 bits are needed. With respect to the traditional binary embedding based methods, Fibonacci embedding based usually does not allow a fixed size embedding since not every pixel of the cover is a "good candidate" for the embedding. To deal with Fibonacci redundancy, it is necessary to comply with Zeckendorf's theorem (see section 3). If the selected pixel is not a "good candidate" (meaning that the current bit to be changed by 1 has a neighbour in the previous bit plane having also a value 1), then the next candidate pixel is selected [1]. In this paper, the redundancy problem of Fibonacci has tolerated without relying on Zeckendorf theorem resulting that each pixel of the cover image can be used for embedding. The most important requirements for steganography systems are payload capacity, stego quality, undetectability, and resistance against active attacks. These requirements cannot be achieved at the same time. Increasing the robustness, generally decrease the invisibility, and increasing the amount of data to be embedded, usually weakens the security (undetectability) and stego quality. The proposed method has the property of high payload capacity (p), embedding 2 bits per pixel, and undetectability, while the stego quality becomes down having said that the PSNR is in the acceptable range of steganography applications, which is above 39 db [2]. Generally, peak signal-to-noise ratio (PSNR) is most commonly used as a measure of the quality of the stego image in field of steganography [3][4]. A larger PSNR value means that the stego image preserves the original cover image quality better. Similar to cryptanalysis, steganalysis attempts to defeat the goal of steganography. However, there have existed many statistical attacks to judge the presence of it and estimate the rate of secret message. Among them, regular and singular (RS) [5], pairs of values (POV) [6], and difference image histogram (DIH) [7] are three steganalyser techniques used to detect and estimate the secret message of the proposed method. These steganalyzers are the most well-known and reliable steganalysers technique. Fortunately, the proposed technique has exceeded all of these three steganalysers (i.e. it is undetectable by them). The rest of the paper is organized as follows. The literature review is presented in section 2. In section 3, the background of Fibonacci is presented. Proposed technique is presented in section 4. Finally, experimental results and conclusion are shown in section 5 and 6 respectively.

## 2. LITERATURE REVIEW

The methods used in image steganography can be grouped into two main categories based on the hiding domain: spatial domain and frequency domain methods. While spatial domain methods use the LSB replacement in bitplanes that represent pixel value such as (binary [8][9][10], Fibonacci [1][11-16], Prime number [17], and Natural number [18] decomposition) sequentially [8][9][10], randomly [10], edge based [19], and etc., frequency domain transforms the image to frequency domain such as DCT [20], wavelet [21] and perform hiding process in that domain. In general, the advantages of Fibonacci embedding based over traditional binary embedding based are the capacity and quality. By providing more planes in Fibonacci, it means more places are available for embedding [11], and by embedding in other than LSB in both binary and Fibonacci, the quality or PSNR of the stego (Fibonacci embedding based) is better than the quality of stego (binary embedding based) [1]. The characteristic of the traditional binary representation of a pixel value is that it is not redundant. This means that the binary decomposition of an integer is unique. While unlike binary representation, the Fibonacci is redundant. This means that more than one sequence can represent the same number. A unique Fibonacci representation is obtained by applying Zeckendorf theorem (discussed in next section). For example, the number 5 can be coded as 1000 or 0110. According to Zeckendorf condition, the code 0110 is not valid.

Diego D. L. Picione and et al [1], they produced an embedding technique based on Fibonacci decomposition. They used Zeckendorf theorem in order to obtain a unique representation for integer number. In their scheme, they first select the pixel then decompose the pixel value into Fibonacci and also select the plane that use for embedding. Then they check the selected pixel if it is a good candidate or not, if it is not, then skip it and next candidate pixel is selected. If it is a good candidate, then the secret bit is replaced with the agreed bit plane. They claimed that the same embedding scheme can be also applied to different planes resulting in more robust data hiding and possibly higher visual distortion. As they mentioned, the main aim of their scheme is to investigate the possibility of inserting a secret bit without altering the perceptual quality of final image (stego). They also claimed that if the secret bits are embedded in the LSB using traditional binary or Fibonacci embedding based, so the PSNR of Fibonacci is similar or higher comparing with the traditional binary embedding based. The limitation of their algorithm is, not every pixel of the cover is going to be used for embedding. In [12], their algorithm's aim is to investigate a different bit planes decomposition based on Fibonacci p-number sequence. They improved the previous scheme in [1] by using generalized Fibonacci decomposition instead the classical Fibonacci. The most common generalization of Fibonacci is the p-number Fibonacci sequence. In order to provide more places for embedding, they investigated the p-number Fibonacci bit planes and see which planes are suitable for embedding. In their scheme, they first decompose the selected pixel into bitplanes using p-number Fibonacci. Then the selected plane is chosen for embedding and also the Zeckendorf theorem is applied on the selected pixel. Finally, they did a comparison between the proposed scheme and classical binary embedding in term of quality and capacity. As a result, even they claimed that the proposed scheme is better than classical binary in term of capacity and quality, but still their limitation is capacity because every pixel of the cover is not able to use for embedding. Furthermore, [11] they produced a new algorithm for embedding that is based on Fibonnaci decomposition. Their algorithm is a modification of the two previous schemes, the classical Fibonacci [1], and the generalized p-number Fibonacci [12]. They modified the Fibonacci sequence by adding any number (non-Fibonacci number) to the sequence, i.e. they convert the Fibonacci sequence into non-Fibonacci sequence. For example, if take a Fibonacci sequence from 1 to 255, this sequence has obtained: 1 2 3 5 8 13 21 34 55 89 144 233, now the Fibonacci decomposition of number 253 is: (1 0 0 0 0 0 1 0 1 0  1 0) comes from 253= 233 + 13 + 5 + 2.   Then by adding any non-Fibonacci number to the list, for example 18, this sequence has obtained: 1  2  3  5  8  13 18  21  34  55  89  144  233 (is called non-Fibonacci sequence). Now the Fibonacci decomposition of number 253 is: (1 0 0 0 0 0 1 0 0 0 0 1 0) comes from 253= 233 + 18 + 2. As they said, by inserting one more number to the sequence, it means one more plane increased into the decomposition and one more place to embed the secret message. Also they claimed that, it is obvious that it can be adding more than one number to the list and that the decomposition is always unique. Once you add any number to the Fibonacci sequence, the produced sequence must have this property:

$$n_1 = 1$$
$$n_1 + n_2 + \ldots + n_k \geq n_{k+1} - 1, \forall\, k \geq 2$$

While $n_i$ is a sequence of non-decreasing positive integer. If this property does not exists, then the produced sequence consider invalid. For example, when number 18 added into the sequence, this sequence has got: 1 2 3 5 8 13 18 21 34 55 89 144 233. Now this sequence is consider valid because, the first element is 1, and $1+2+3+5+8+13 \geq 18-1$.

In their scheme, the selected pixel is decomposed, and then replaces the secret bit with the agreed plane. They claimed that the advantage of their method over other Fibonacci based methods is the fact that given a non-Fibonacci list of integers, may circumvent one of the problems that the original method has, that of not being able to use some pixels. Since the original method uses the Zeckendorf's theorem, in the decomposition, we cannot have two consecutive aces. This means that we cannot use some pixels, as by embedding the data we might end up with a non-Zeckendorf compatible decomposition. They also claimed by using another list of integers, a complete sequence, we enable ourselves to bypass the limitations of Zeckendorfs decomposition. In [13], they produced a new version of Fibonacci. In their algorithm, they modified the generalized Fibonacci sequence [12] by adding two parameters p, r. These two parameters increase the security of the whole system; without their knowledge it is not possible to perform the same decomposition used in the embedding process and to extract the embedded information. They claimed that the classical binary embedding technique is equivalent (similar) to Fibonacci based embedding in term of PSNR if the embedding is performed in LSB. While, when the embedding is performed in rather than LSB, the PSNR of Fibonacci is better than classical binary. Therefore they suggested that in Fibonacci, different bitplanes other than LSB can be used for embedding. As they claimed in their conclusion, their algorithm has the advantage in term of security (because of p, r) and quality comparing with other previous Fibonacci and binary techniques. It is notice that each element in the sequence is obtained by adding the previous r elements taken at distance p, and it is necessary to fulfill the following constraints:

- A valid (p, r) Fibonacci coefficient vector c must contain less than p-1 zeros between two ones.
- A valid (p, r) Fibonacci coefficient vector c cannot contain more than r consecutive groups, being constituted by one symbol equal to 1 followed by p-1 symbols equal to 0.

It can be noticed that when p=0, we obtain the classical binary sequence, and when p=1, we obtain the classical Fibonacci sequence. In [13], their algorithm has a problem of capacity, because not every pixel of the cover is a good candidate to use for embedding relying on Zeckendorf theorem. In [14], they produced an embedding technique based on Fibonacci sequence. They embed secret message in Fibonacci cover. They claimed that the main aim of their algorithm is to make their system more robust against detection. They do embedding in areas with high activity such as contours, texture, and eventually noise, which are more robust, while they avoid embedding in low activity areas such as flat regions. For detecting high activity and low activity areas, they used LAI (Local Activity Index). They do inserting the secret bits in LSB of the Fibonacci code. They said, if all blocks have been utilized, but the secret message has not been completely hidden in the cover, the next plane of the Fibonacci domain is selected for embedding. They claimed that by increasing p and r the uniqueness become more restrictive and blocks with lower activity and higher bit-planes are going to be used. They also claimed that their proposed algorithm has better result than classical binary decomposition in both security and perceptual aspects. Again, this algorithm has a limitation of capacity such that not every pixel is used for embedding. In [15], the main aim of their algorithm is capacity and security. They proposed a new adaptive steganography system based on Fibonacci. Their algorithm is based on Fibonacci in order to gain more allowable embedding capacity. They also use T- order statistics in order to make their algorithm more resistance to detection against various steganalysis tools. T-order statistics enables the embedding of secret data only in the noisy regions making any changes to the cover undetectable. They claimed that good locations (noisy regions) for embedding are selected based on local information gathered in the vicinity of a given cover pixel. They also claimed that many approaches have been proposed using such local measures as variance, standard deviation, median-based variance, as well as the number of unique pixel values within a given distance. They said that if the calculated variation for a particular pixel exceeds the threshold by a determined magnitude, the pixel may be used for the insertion of multiple bits; therefore this has an effect of increasing the embedding capacity. In [16], their algorithm is based on Fibonacci embedding. They claimed that their proposed algorithm proved to have a greater resistance to detection as well as an increase in embedding capacity with existing adaptive and non-adaptive methods. They applied variation measures to select noise area that used for embedding i.e. the embedding process is first select the noise area then decomposed pixel into Fibonacci. In [15][16], they claimed that their algorithm has a property of capacity, because there are more bitplanes

(12 bitplanes) are available for embedding, but still they have the capacity limitation because cannot be use every pixel for embedding. The proposed method has tolerated the redundancy problem of Fibonacci and without relying on Zeckendorf theorem resulting that each pixel of the cover image can be used for embedding.

## 3. BACKGROUND

One of the most famous integer sequences is the Fibonacci sequence. The classical Fibonacci number was introduced in 13th century by Leonardo of Pisa [11]. The sequence is given by the following recursive formula:

$$F_n = \begin{cases} 1 & n = 0 \\ 1 & n = 1 \\ F_{n-1} + F_{n-2} & n > 1 \end{cases} \quad (1)$$

This sequence is a particular case of a larger family of sequences. The most common generalization of Fibonacci numbers is the Fibonacci p-number sequence defined as follows [12]:

$$F_n = \begin{cases} 1 & n = 0 \\ 1 & n = 1 \\ F_{p(n-1)} + F_{p(n-p-1)} & n > 1 \end{cases} \quad (2)$$

In which each element depends on the previous one and on the p-th previous element of the sequence. In fact each decimal number can have more than one representation. To obtain a unique representation of a number, the mathematician Zeckendorf proved the following theorem: "Every positive integer can be uniquely represented as a sum of non-consecutive Fibonacci numbers" [11]. For example, if we want to find the Fibonacci sequence of number (20) with p=1, r=1: The Fibonacci sequence is: 1 1 2 3 5 8 13 21. Then the Fibonacci sequence of number 20 is: 20 = 13 + 5 + 2, and the Fibonacci decomposition of 20 is: 1 0 1 0 1 0. Put 1 for each number that exists in the Fibonacci representation, and 0 for those are not exist, then by taking the inverse of the Fibonacci code sequence. Figure 1(A, and B), are Fibonacci sequence and Fibonacci code for the number 255. As we know that the pixel value is non-negative integer less than 256, this means that it can be written in binary form using 8 bits; similarly, we can write the pixel value in Fibonacci form using 12 bits.

| A. Generalized Fibonacci sequence for number 255 |||||||||||||
|---|---|---|---|---|---|---|---|---|---|---|---|
| 1 | 2 | 3 | 5 | 8 | 13 | 21 | 34 | 55 | 89 | 144 | 233 |
| B. Fibonacci code to represent number 255 |||||||||||||
| 1 | 0 | 0 | 0 | 0 | 0 | 1 | 0 | 0 | 0 | 0 | 1 |

Figure 1. Fibonacci sequence and Fibonacci code

## 4. PROPOSED TECHNIQUE

Once the pixel value converts into Fibonacci bit-planes, there should be each pixel value has more than one representing. For this sake, the pixel value must convert to Fibonacci according to the Zeckendorf theorem i.e. there is no two consecutive ones in the Fibonacci. According to this theorem, the probabilities of first three LSBs of a cover pixel in Fibonacci representation are (000, 001, 010, 100, 101). And by taking two bits at a time of secret message in binary representation we have (00, 01, 10, 11). Depending on these probabilities, our mapping is proposed as illustrated in 4.1.

### 4.1 Proposed mapping algorithm

By embedding the secret message into cover (Fibonacci based) according to the proposed mapping algorithm that illustrated in table 1. The embedding process is used the first three LSBs of the cover.

Table 1. Proposed mapping algorithm

| Cover bits \ Secret bits | 00 | 01 | 10 | 11 |
|---|---|---|---|---|
| 000 | 000 | 001 | 100 | 101 |
| 001 | 000 | 001 | 100 | 101 |
| 010 | 010 | 001 | 100 | 101 |
| 100 | 000 | 001 | 100 | 101 |
| 101 | 000 | 001 | 100 | 101 |

By extracting from the first and third LSBs of the selected pixel of the stego, we get the secret message. It is important to notice that when we have this Fibonacci code: 0 0 0 0 1 0 0 1 0 0 1 and we have to map this two secret bits 11, according to the mapping algorithm in table 1, the first three bits of the cover becomes 101 (i.e. the Fibonacci code becomes: 0 0 0 0 1 0 0 1 1 0 1 ) and this Fibonacci code is not valid according to Zeckendorf theorem, because we have two consecutive ones. For solving this problem, we have to put a condition, if the last bit of the three selected bits of the cover is zero (for example 001) and the secret bit to be embedding is 1, and also the previous neighbor bitplane of the selected three bitplanes of the cover, i.e. 4th LSB, is 1 (i.e. 1 001), we need to change the previous neighbor by zero then embed the secret bit such (0 101).

## 5. EXPERIMENTAL RESULTS

To evaluate the proposed technique, two experiments are performed. One is for the capacity (discussed in 5.1) and the other is for detectability (discussed in 5.2). Then the results are compared with the three steganographic techniques (LSB sequential [8][9][10], LSB randomly [10], and Fibonacci- randomly embedding [1]). For testing each algorithm, five different cover-images (see figure 3) with size 512 x 512 are used. The secret bits have been generated using the Matlab PRNG ( Psudo Random Number Generator). For all tested techniques (except the proposed), for each cover image, we have generated four stego images by embedding four different message length p = 0.25, 0.5, 0.75, and 1 corresponding to %25, %50, %75, and %100 of the total pixel number of the cover-image respectively, while for the proposed technique the message length p = 0.25, 0.5, 0.75, 1, 1.5, and 2 corresponding to %25, %50, %75, %100, %150, and %200 of the total pixel number of the cover-image (i.e. six stego images have been produced). After embedding, for each case, i.e. for each technique with a specific embedding rate, 5 stegos are produced; the average results of PSNR between the covers and stegos for each technique are illustrated in figure 2. This shows that the all embedding techniques (except the proposed) have the highest PSNR. This result can be attributed to the fact that for these methods only LSB has changed, while in the proposed technique the first 3 LSBs (some time first 4 LSBs) are changed. It is notice that when the secret length p = %100, the PSNR of traditional Fibonacci is higher than the others, because some of the secret bits are not going to be embed as some cover pixels are skip for embedding according to the Zeckendorf theorem. Furthermore, this makes the capacity ratio of traditional Fibonacci become reduce (see table 2 in 5.1).

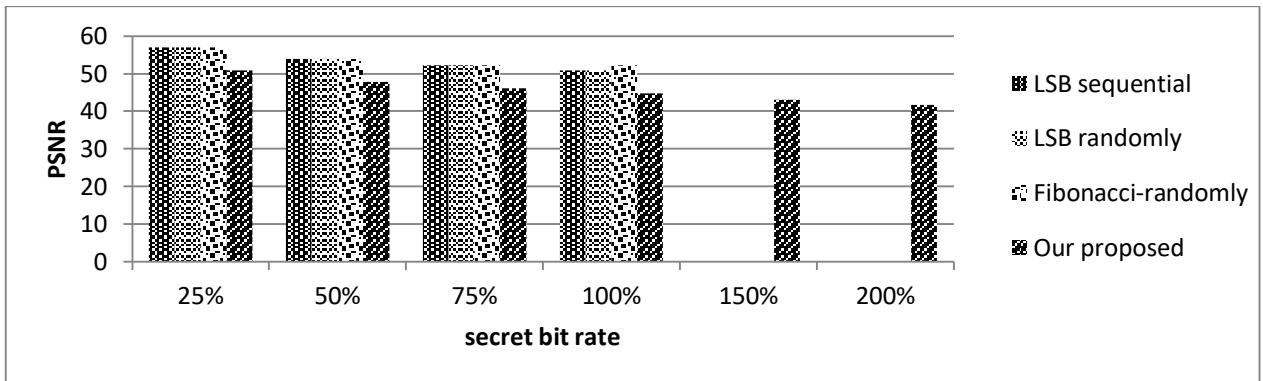

Figure 2.Stego quality

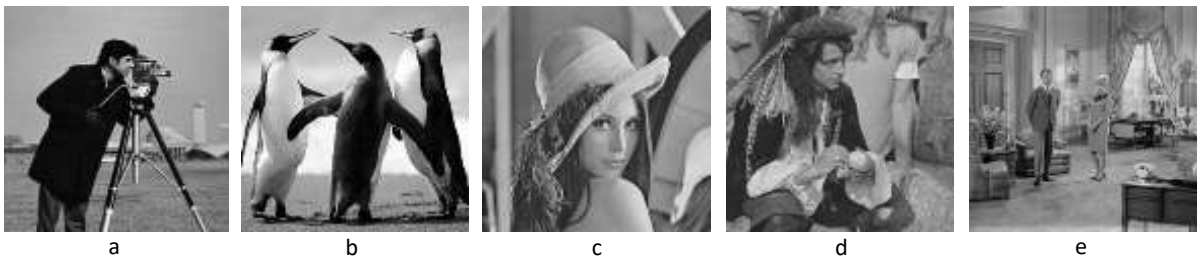

a  b  c  d  e

Figure 3.Cover images

## 5.1 Capacity:

Because the cover images sizes are 512 x 512, therefore the maximum numbers of provided pixels for embedding are 262144 pixels. In the table 2, the capacity of all techniques including the proposed are illustrated with different ratio (0.25, 0.50, 0.75, 1.0, 1.5, and 2.0). From the table 2, you can notice that the maximum capacity ratio for tested techniques is 1.0, and it is 2.0 for our proposed corresponding to %100 and %200 respectively. It also noticeable that in the case of capacity ratio = 0.75, and 1.0, the capacity of Fibonacci-randomly is 194189 bits, and 196273 bits respectively, which are less than 196608 bits, and 262144 bits (which are the maximum number of pixels per bit). This was happened because in the case of Fibonacci-randomly, some pixels are going to be skipped according to Zeckendorf theorem. Therefore, we can say that although the stego quality becomes down within acceptable range (as illustrated in figure 2), the proposed technique has the capacity double of binary based techniques (sequential, and randomly), and has more than double capacity over Fibonacci-randomly. In each different ratio, the number of pixels that used for embedding in our proposed is half of the number of pixels that used in other techniques.

Table 2. Capacity rate

| Capacity rate | 0.25 | 0.50 | 0.75 | 1.0 | 1.5 | 2.0 |
|---|---|---|---|---|---|---|
| LSB sequential | 65536 | 131072 | 196608 | 262144 | | |
| LSB randomly | 65536 | 131072 | 196608 | 262144 | | |
| Fibonacci- randomly | 65536 | 131072 | 194189 | 196273 | | |
| Proposed | 65536 | 131072 | 196608 | 262144 | 393216 | 524288 |

## 5.2 Detectability and message estimation:

In this section, three steganalysers techniques have been used to test the detectability of the proposed technique and then compare it with the other techniques mentioned in section 5.

### 5.2.1 RS steganalyser:

The method defines two functions: $F_1$, which changes a pixel value $0 \leftrightarrow 1, 2 \leftrightarrow 3, 4 \leftrightarrow 5, ...,254 \leftrightarrow 255$ and $F_{-1}$, which changes a pixel value $-1 \leftrightarrow 0, 1 \leftrightarrow 2, 3 \leftrightarrow 4, ...,255 \leftrightarrow 256$. RM is the ratio of the blocks in which the total of fluctuations increases when $F_1$ is applied to the blocks with mask M. SM is the ratio of blocks in which the total of fluctuations decreases when $F_1$ is applied to the blocks with mask M. Also, RM- and SM- are defined with $F_{-1}$, instead of $F_1$. Fridrich found that the RS ratio of a typical image should satisfy the rule: RM $\cong$ RM- and SM $\cong$ SM- through large amount of experiments. When only LSB of the original cover is changed, the difference between RM and RM- and the difference between SM and SM- increase. Then, the rule is violated; therefore, one could conclude that the tested image has a hidden message. Depending on the description above, we can discuss table 3. Each cell is representing the average result of five tested images. When p = % 0, i.e. images without embedding, the value of RM is close to RM-, and also the value of SM is close to SM-. In the case of using LSB sequential and LSB randomly techniques, by increasing the embedding rate, i.e. p, the differences between RM and RM-, SM and SM- are increased. This indicates that these images are containing secret message. While in the case of using Fibonacci-randomly and the proposed embedding techniques, there are not such differences. This indicates that these images are non-stego images. Therefore, the proposed technique is exceeded the RS steganalyser.

Table 3. RS steganalyser

| | P | % 0 | % 25 | % 50 | % 75 | % 100 | % 150 | % 200 |
|---|---|---|---|---|---|---|---|---|
| LSB sequential | RM- | 42.3217773 | 45.4031372 | 47.7227783 | 49.8452758 | 51.676635 | | |
| | RM | 41.0702514 | 37.6742553 | 34.9624633 | 32.6904296 | 30.9039306 | | |
| | SM- | 21.6735839 | 20.3948974 | 19.4607543 | 18.6499023 | 17.5546264 | | |
| | SM | 22.8009033 | 25.5178833 | 27.7090454 | 29.4885253 | 31.0055542 | | |
| LSB randomly | RM- | 42.3217773 | 45.1400756 | 47.713012 | 49.7644043 | 51.7785644 | | |
| | RM | 41.0702514 | 38.4375 | 35.8502193 | 33.3828735 | 30.9976196 | | |

| | P | % 0 | % 25 | % 50 | % 75 | % 100 | % 150 | % 200 |
|---|---|---|---|---|---|---|---|---|
| | SM- | 21.6735839 | 20.5453491 | 19.4647216 | 18.460083 | 17.5842285 | | |
| | SM | 22.8009033 | 24.628601 | 26.654663 | 28.673095 | 30.9371948 | | |
| Fibonacci-Randomly | RM- | 42.3217773 | 42.1401977 | 41.7730712 | 41.2393188 | 41.1953735 | | |
| | RM | 41.0702514 | 41.0467529 | 40.8523559 | 40.4281616 | 40.4208374 | | |
| | SM- | 21.6735839 | 21.9512939 | 22.3025512 | 22.5003051 | 22.517395 | | |
| | SM | 22.8009033 | 23.0316162 | 23.2220459 | 23.389587 | 23.395996 | | |
| Proposed | RM- | 42.3217773 | 42.0251464 | 41.6705322 | 41.446533 | 41.0769653 | 40.3060913 | 39.8373413 |
| | RM | 41.0702514 | 40.8630371 | 40.7254028 | 40.671997 | 40.5914306 | 40.5398559 | 40.298767 |
| | SM- | 21.6735839 | 22.5863647 | 23.4387207 | 24.3075561 | 25.0314331 | 26.2905883 | 27.2781372 |
| | SM | 22.8009033 | 23.572387 | 24.1793823 | 24.8831176 | 25.3192138 | 26.2179565 | 26.9177246 |

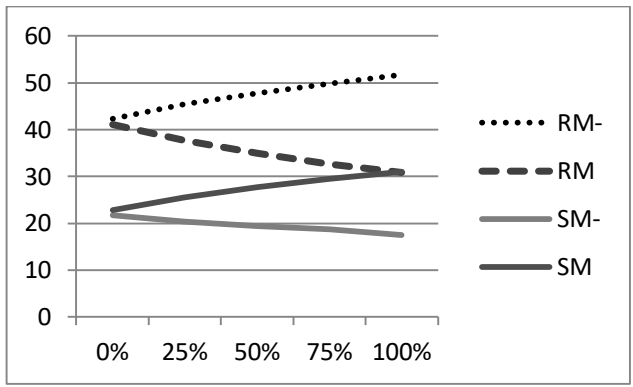

A

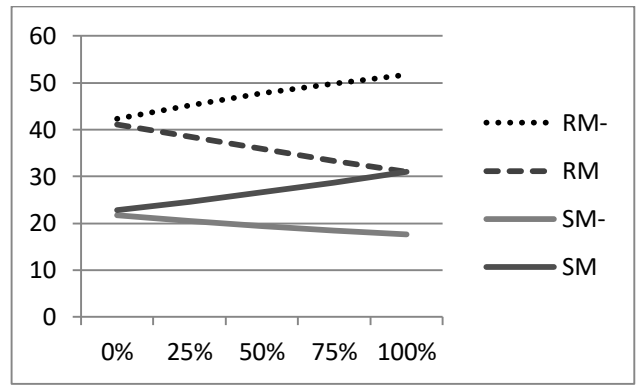

B

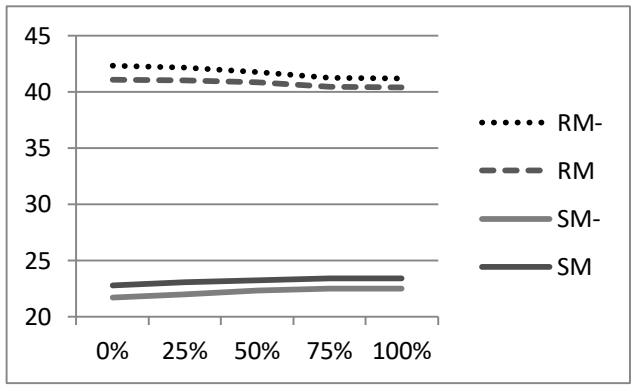

C

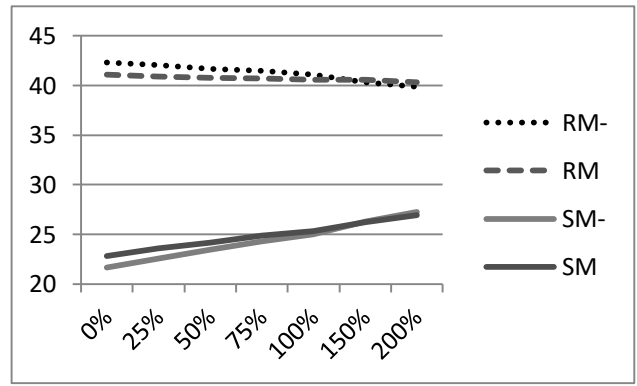

D

Figure 4. RS diagrams for LSB sequential (A), LSB randomly (B), Fibonacci-randomly (C), and proposed method (D). The x-axis is the relative number (ratio) of pixels with flipped LSBs, the y-axis is the relative number of regular and singular groups (RM,RM-,SM,SM-).

### 5.2.2 POV steganalyser:

The test makes the statistical probability of embedding using Chi-square test. If the steganalyst is presented with a plot similar to that in figure 5 (A), then the image should assume has not been manipulated, while if the plot is like (B) and (C) in figure 5, then the image is assumed that containing % 50 and % 100 embedding capacity respectively. For the convenience of display, the result of only three images, a, b, and c (see figure 3), out of five images are displayed, and only for % 100 embedding capacity (full capacity) is displayed except that for the proposed technique for %200 embedding capacity is displayed too. From figures (10, and 11), we can notice that almost all plots of our proposed are exactly like figure 5 (A) which means that they are considered as a non-stego images.

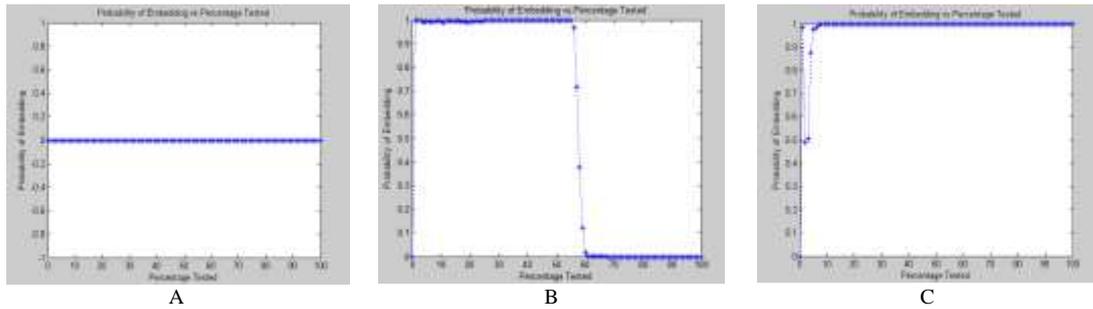

A　　　　　　　　　　　　　B　　　　　　　　　　　　　C
Figure 5. Some instance of POV plots: (A) mean zero embedding, (B) mean %50 embedding, and (C) mean %100 embedding.

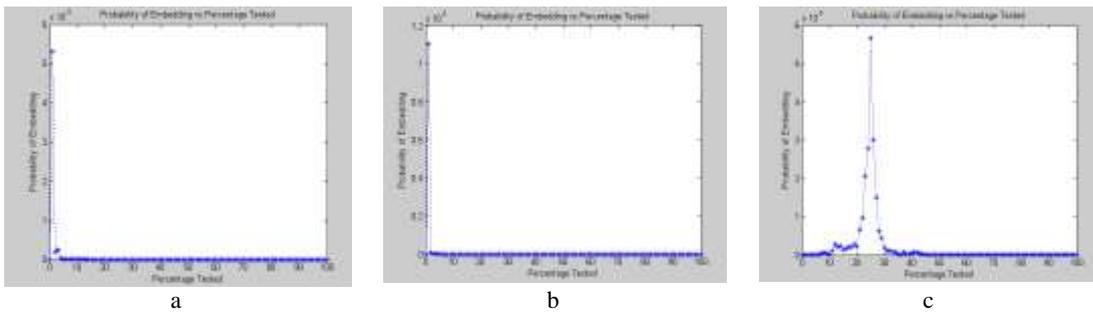

a　　　　　　　　　　　　　b　　　　　　　　　　　　　c
Figure 6. POV of cover-images

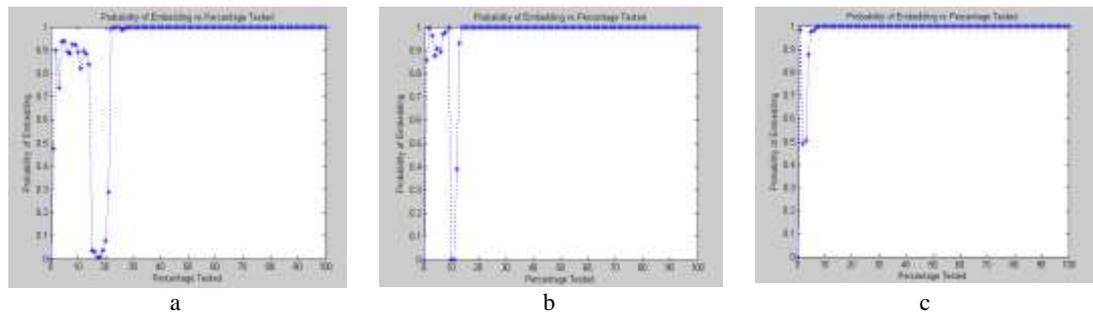

a　　　　　　　　　　　　　b　　　　　　　　　　　　　c
Figure 7. POV of LSB sequential (%100) embedding

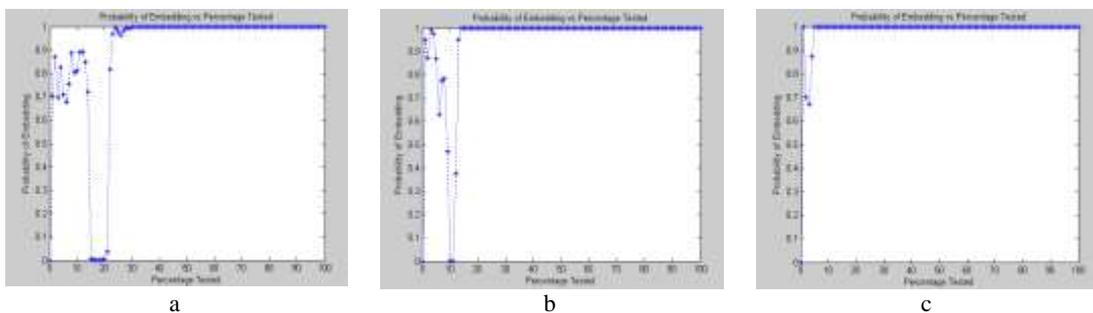

a　　　　　　　　　　　　　b　　　　　　　　　　　　　c
Figure 8. POV of LSB randomly (%100) embedding

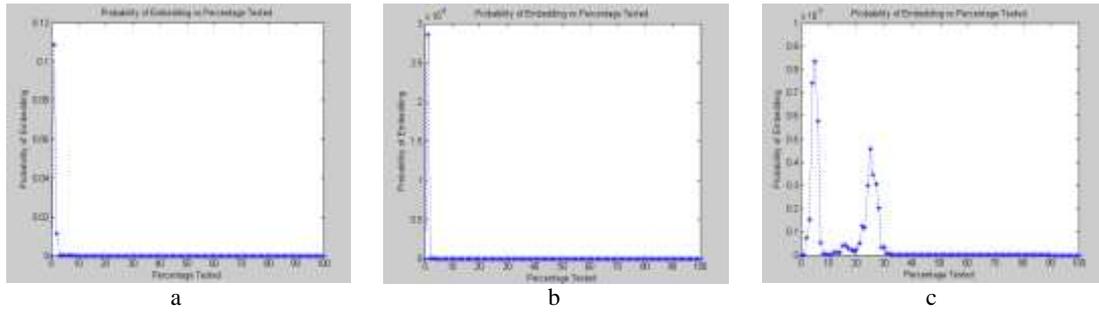
a  b  c
Figure 9. POV of Fibonacci-randomly (%100) embedding

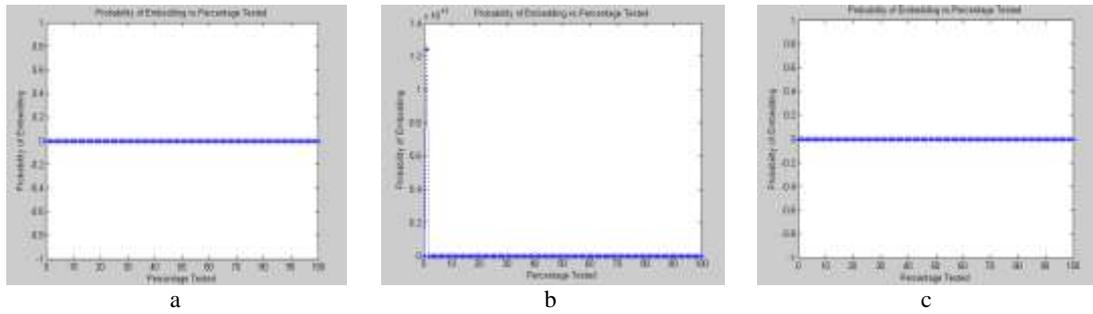
a  b  c
Figure 10. POV of proposed (%100) embedding

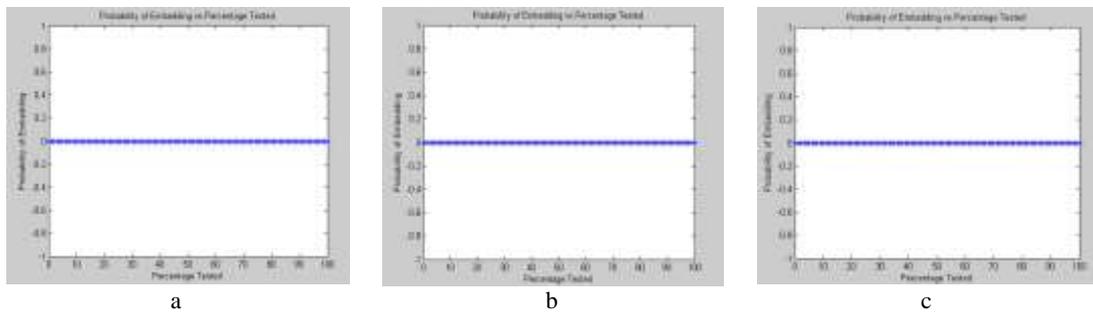
a  b  c
Figure 11. POV of proposed (%200) embedding

### 5.2.3 DIH Steganalyser:

It is another reliable steganalyser technique that uses the measure of weak correlation between the LSB plane and its neighbor bitplanes to construct a classifier for discrimination between stego-image and cover-image. The technique can obtain the embedded message ratio as illustrated in table 4. From the table below, it is noticeable that our proposed has exceeded the message estimation ratio.

Table 4. DIH steganalyser

|  | P | % 0 | % 25 | % 50 | % 75 | % 100 | % 150 | % 200 |
|---|---|---|---|---|---|---|---|---|
| LSB sequential | Image a | -0.033596 | 0.31370304 | 0.64905220 | 0.8677309 | 0.8941849 | | |
| | Image b | 0.1213431 | 0.57800796 | 0.72867644 | 0.76949980 | 0.8602725 | | |
| | Image c | 0.100168 | 0.28601087 | 0.49171606 | 0.76339162 | 0.9064198 | | |
| | Image d | 0.0837682 | 0.37976808 | 0.47101877 | 0.6679757 | 0.8091537 | | |
| | Image e | 0.1133309 | 0.27217528 | 0.49493741 | 0.7156911 | 0.9045631 | | |
| LSB randomly | Image a | -0.033596 | 0.26124636 | 0.5494377 | 0.8476524 | 0.9587756 | | |
| | Image b | 0.1213431 | 0.35948136 | 0.58910637 | 0.7821523 | 0.9720207 | | |
| | Image c | 0.100168 | 0.37384542 | 0.5973690 | 0.7682262 | 0.8823120 | | |

|  | P | % 0 | % 25 | % 50 | % 75 | % 100 | % 150 | % 200 |
|---|---|---|---|---|---|---|---|---|
|  | Image d | 0.0837682 | 0.36285485 | 0.50449975 | 0.7210255 | 0.8807905 |  |  |
|  | Image e | 0.1133309 | 0.29169098 | 0.62812427 | 0.8099179 | 0.9371658 |  |  |
| Fibonacci-randomly | Image a | -0.033596 | -0.0518723 | 0.0200270 | 0.0054271 | 0.0143204 |  |  |
|  | Image b | 0.1213431 | 0.20225848 | 0.20821765 | 0.1888501 | 0.1888501 |  |  |
|  | Image c | 0.100168 | 0.09770634 | 0.12807834 | 0.1776499 | 0.1733854 |  |  |
|  | Image d | 0.0837682 | 0.13735064 | 0.06654268 | 0.0254803 | 0.0254803 |  |  |
|  | Image e | 0.1133309 | -0.0133655 | -0.0790132 | -0.1681111 | -0.169226 |  |  |
| Proposed | Image a | -0.0335960 | -0.0084963 | 0.00186904 | 0.0204315 | 0.009868 | 0.0278307 | 0.0645702 |
|  | Image b | 0.1213431 | 0.09635262 | 0.13789408 | 0.1130857 | 0.0797286 | -0.0179749 | -0.0600332 |
|  | Image c | 0.100168 | 0.05178603 | 0.05069195 | 0.0554710 | 0.0621517 | 0.0249011 | -0.0830512 |
|  | Image d | 0.0837682 | 0.02553571 | 0.02241638 | 0.0071408 | 0.0082813 | -0.1273610 | -0.1349887 |
|  | Image e | 0.1133309 | 0.10089804 | 0.10642381 | 0.10487022 | 0.0891117 | 0.0211184 | -0.0149882 |

## 6. COCLUSION

Fibonacci based embedding technique has been proposed to achieve efficient steganography technique in terms of capacity. Innovative idea that extends Fibonacci-like steganography by bit-plane(s) mapping instead of bit-plane(s) replacement has been investigated. Our proposed algorithm increases embedding capacity using bit-plane mapping that embeds two bits of the secret message in three bits of a pixel of the cover, at the expense of reasonable loss in stego quality. While existing Fibonacci embedding algorithms do not use certain intensities of the cover for embedding due to the limitation imposed by the Zeckendorf theorem, our proposal solve this problem and make all intensity values candidates for embedding. Our experimental results indicate that the proposed technique double the embedding capacity when compared to existing Fibonacci methods, and it has robustness advantage against some statistical attacks such as RS, POV, and DIH.